\documentclass[aps,prl,twocolumn,superscriptaddress,showpacs]{revtex4}

\usepackage{longtable}
\usepackage{graphicx}
\usepackage{dcolumn}
\usepackage{bm}
\usepackage{amsmath}
\usepackage[psamsfonts]{amssymb}
\usepackage{subfigure}

\newcommand{\PCCO}{Pr$_{2-x}$Ce$_x$CuO$_{4-\delta}$}

\newcommand{\AFM}{antiferromagnetic}

\newcommand{\etal}{\emph{et al.}}

\begin{document}

\title{Evolution of a bosonic mode across the superconducting dome in the high-T$_{c}$ cuprate Pr$_{2-x}$Ce$_x$CuO$_{4-\delta}$}


\author{I. Diamant}
\email[]{diamanti@post.tau.ac.il}\affiliation{Raymond and Beverly Sackler School of Physics and Astronomy, Tel-Aviv
University, Tel Aviv, 69978, Israel}
\author{S. Hacohen-Gourgy}
\affiliation{Raymond and Beverly Sackler School of Physics and Astronomy, Tel-Aviv University, Tel Aviv, 69978, Israel}
\author{Y. Dagan}
\affiliation{Raymond and Beverly Sackler School of Physics and Astronomy, Tel-Aviv University, Tel Aviv, 69978,
Israel}


\date{\today}

\begin{abstract}
We report a detailed spectroscopic study of the electron doped cuprate
superconductor Pr$_{2-x}$Ce$_x$CuO$_{4-\delta}$ using point
contact junctions for x=0.125(underdoped), x=0.15(optimally doped) and
x=0.17(overdoped).
From our conductance measurements we are able to identify bosonic
resonances for each doping. These excitations
disappear above the critical temperature, and above the critical
magnetic field. We find that the energy of the bosonic excitations
decreases with doping, which excludes lattice vibrations as the paring glue. We conclude that the bosonic mediator for these cuprates is more likely to be
spin excitations.
\end{abstract}


\pacs{74.50.+r, 74.62.Dh, 74.72.-h}

\maketitle
In the standard theory for superconductivity by Bardeen, Cooper and Schrieffer \cite{BCS}
electrons pair and condense with a characteristic energy scale
$\Delta$. This scale manifests itself as  a gap in the density of
states which can be probed by tunneling spectroscopy.\cite{Giaever} The pairing between electrons occurs despite the
coulomb repulsion through a bosonic mediator - "the
glue". 
In their hallmark work McMillan and  Rowell\cite{McmillanPhonon} showed that for
strong enough electron-phonon coupling the tunneling spectra
should deviate from the simple BCS density of states.
In their work they studied the Lead phonon  spectrum and unequivocally established
them as the paring glue in the conventional BCS\cite{BCS}
superconductors. 
%
\par
Electron pairing also occurs in the high temperature cuprate
superconductors.\cite{GoughElectronPairs}. A common characteristic to all these
superconductors is the strong dependence of their electronic
properties on the number of charge carriers put into the cooper
oxygen planes (doping). These charge carriers can be either holes
(p-doped) or electrons (n-doped). Many experiments focused on the
hole-doped cuprates, while the electron-doped ones remained less
studied. Despite many years of research, both on the hole and electron doped side of the phase diagram, there is still a question
mark on the nature of the pairing glue. Several theoretical
possibilities were suggested: pairing by magnetic
interactions,\cite{MonthouxSpinFluctuations, MonthouxPinesSpinFluctuations, ChubukovSpinFluctuationModel, ManskeSpinTheory} phonons \cite{SongPhononCoupling}, and pairing without
a glue.\cite{AndersonGlue}

A number of experiments found evidence for bosonic modes in the
hole-doped cuprates. However, they could not come to an agreement
on the nature of the pairing bososn. Pairing by lattice vibrations
was suggested by some of them
\cite{LanzaraPhononCoupling,LeeBosonicMode,McQueeneyPhononCoupling}
while others were interpreted in terms of paring by spin resonances.\cite {JenkinsFischerSpinExcitation, NormanSpinCoupling,MignodSpinCoupling} 
\par
On the electron doped side of the phase diagram it has been recently reported that bosonic modes can be found in optimally doped Pr$_{0.88}$LaCe$_{0.12}$CuO$_4$ (PLCCO). \cite{NiestemskiBosonicMode} The deduced energy scale is found to be consistent with both acoustic phonons and spin excitations. 
\par
Neutron scattering experiments on optimally doped PLCCO were able to associate a magnetic resonance to
superconductivity and to scale the resonance energy with the
critical temperature by: $E_r\approx5.8 k_BT_c$, as was found for
other hole-doped cuprates.\cite{WilsonNeutron}
\par
Here we show that using point spectroscopy on the n-doped cuprate \PCCO~ (PCCO) we are able to
identify bosinc modes and follow their doping dependence both
on the overdoped and the underdoped sides of the phase diagram. From this
dependence we conclude that the bosons responsible for the pairing
interaction in the electron-doped cuprates are most probably spin
excitations.
\par
We used pulsed laser deposition to fabricate thin films of the
electron-doped cuprate superconductor
Pr$_{2-x}$Ce$_x$CuO$_{4-\delta}$. Three doping levels were
studied: x=0.125 (underdoped), x=0.15 (optimally doped), and
x=0.17 (overdoped). The sample was approached by a a micro edge
Pt-Ir alloy tip using a point contact probe with a 15nm step size.
This setup enabled us to vary the height of the potential barrier
by changing the sample-tip distance. The distance was varied until
clear superconducting features were seen. 

\par
\begin{figure*}
\begin{center}
\subfigure[Underdoped x=0125]{\label{fig:0125}\includegraphics[width=.25\textheight]{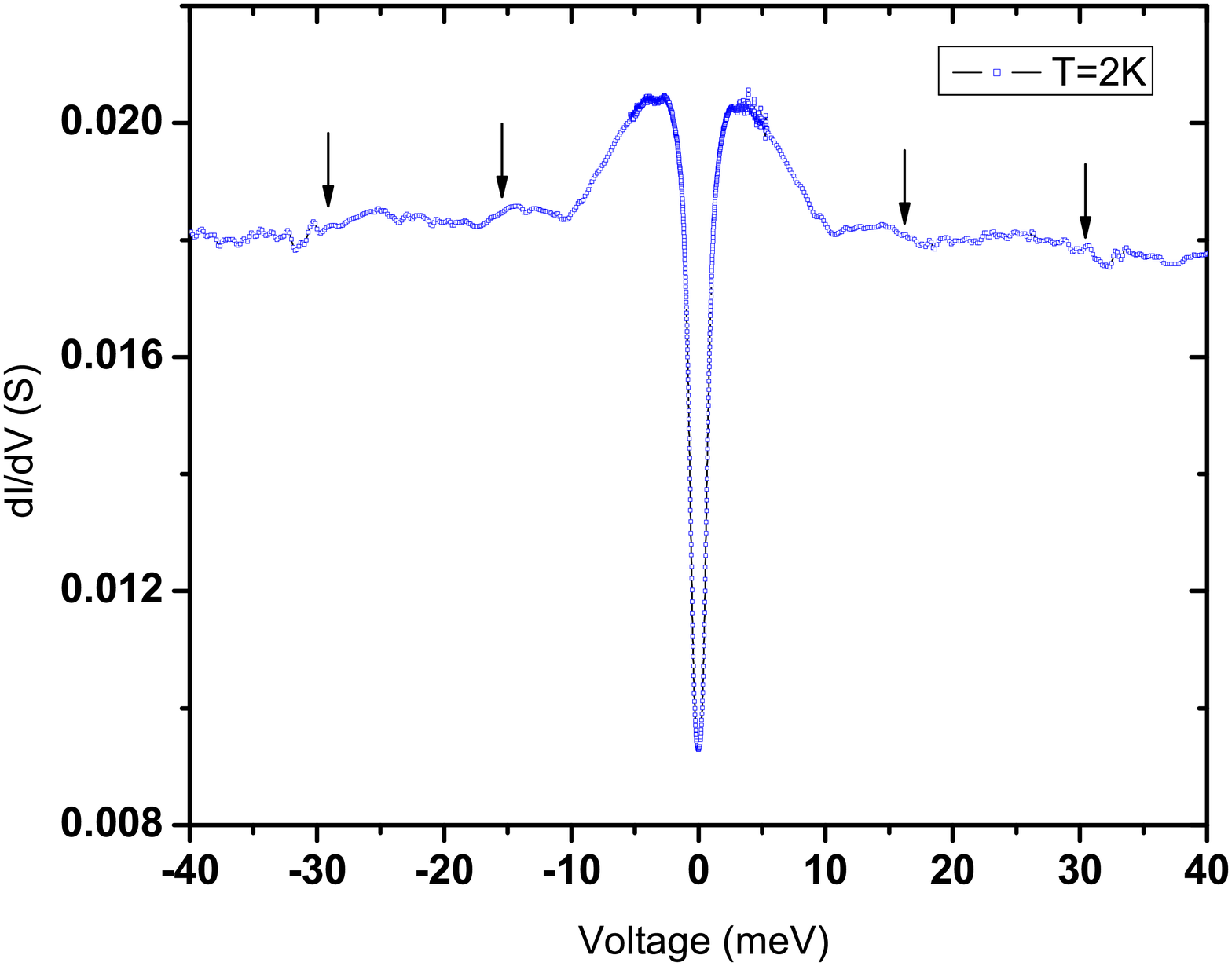}}
\subfigure[Optimally doped x=015]{\label{fig:015}\includegraphics[width=.25\textheight]{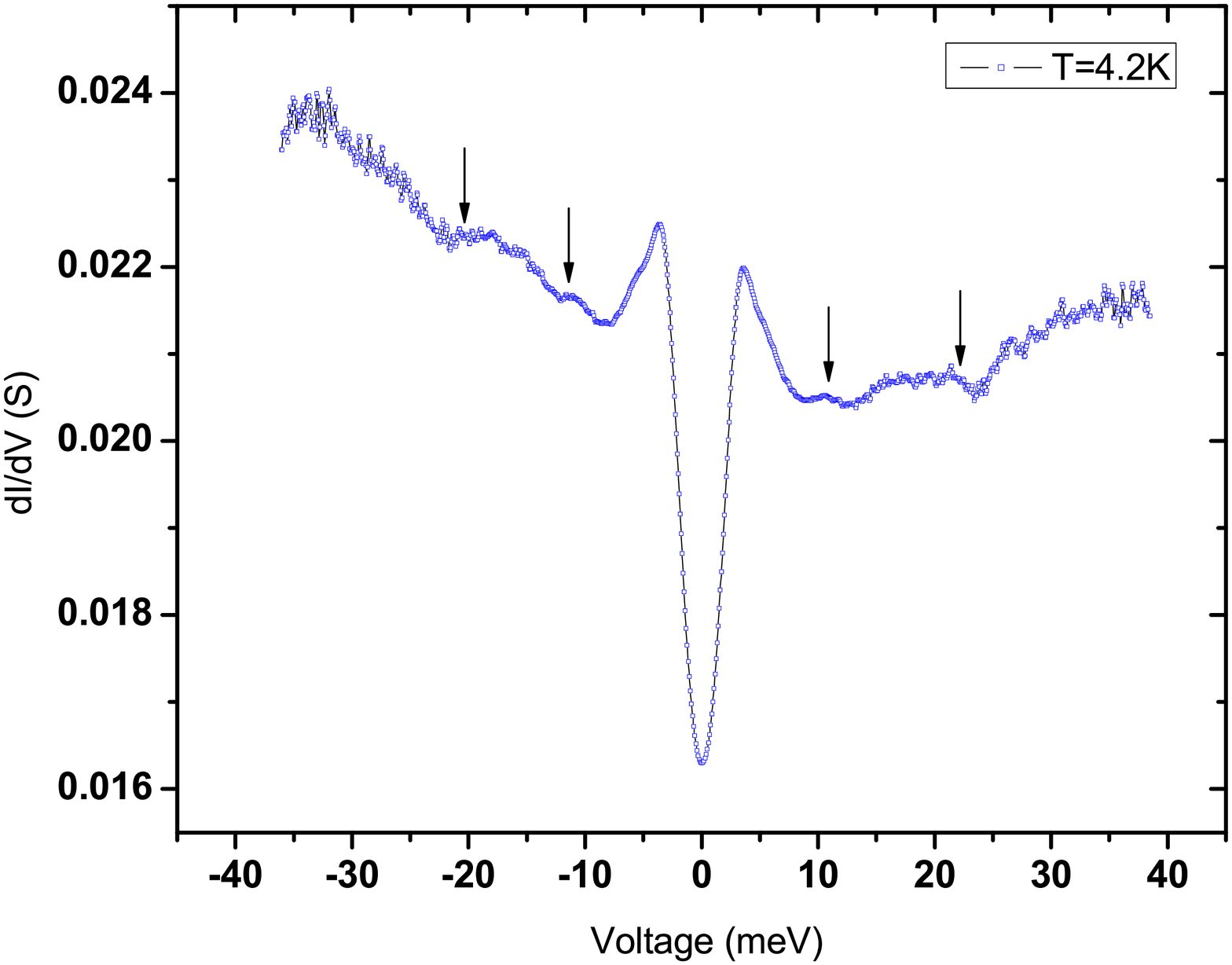}}
\subfigure[Overdoped x=017]{\label{fig:017}\includegraphics[width=.25\textheight]{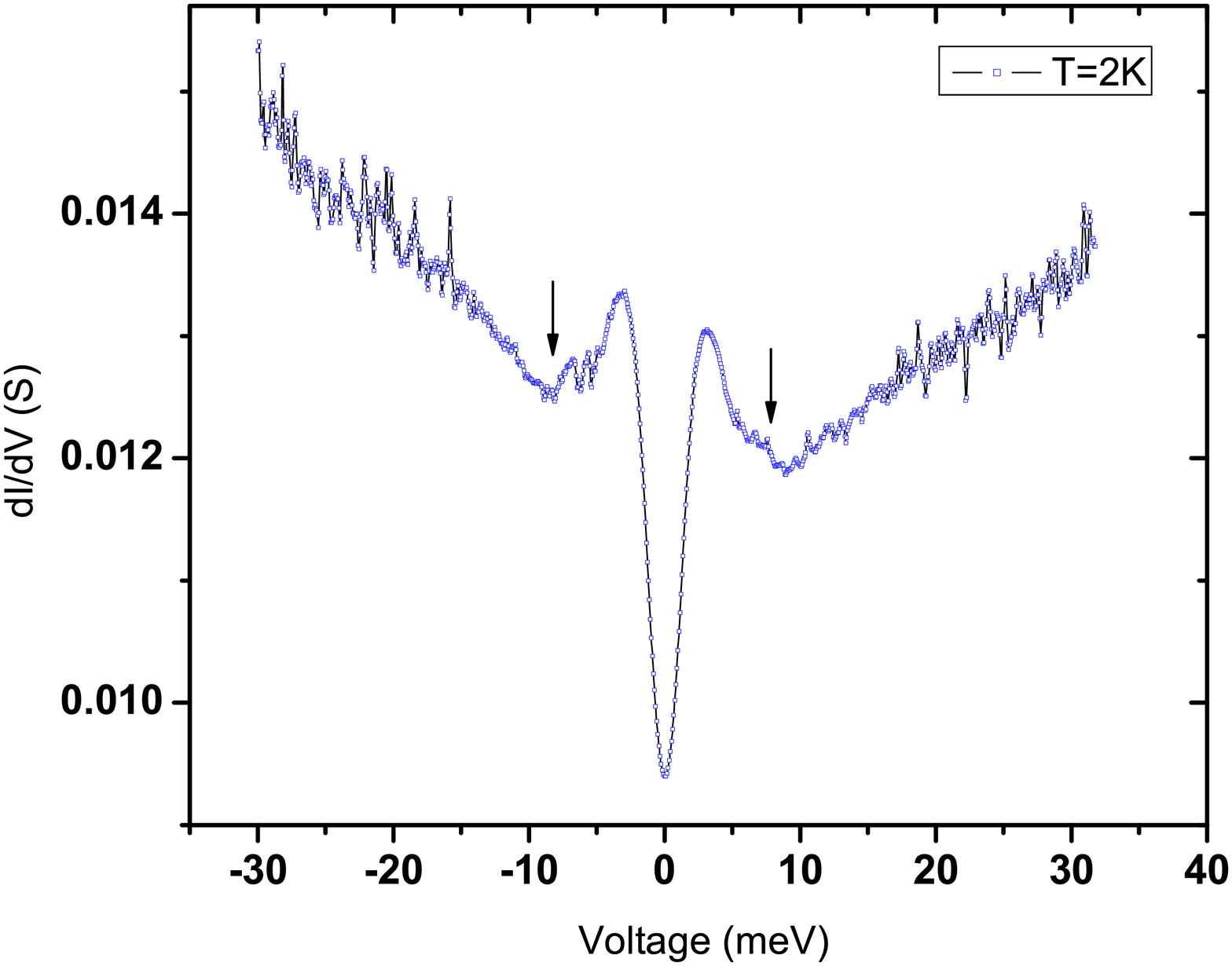}}

\caption {(Color online) The differential conductance as a function of the applied bias at low temperature. We present three doping levels: (a) the extremely underdoped $x=0.125$ sample: J0105, (b) the optimally doped $x=0.15$ Sample: J0803, and (c) the overdoped $x=0.17$ sample: J0224. The superconducting energy gap $\Delta$ is determined by the location
of the coherence peaks. The obtained values are consistent with
previous measurements.\cite{DiamantSingle} Additional features
appear at energies higher than $\Delta$. These features are absent
above the critical temperatures, and
above the critical magnetic field. The parameters of each junction are: For extremely underdoped sample $x=0.125$ (a) the gap amplitude is $\Delta=2.3\pm0.3meV$, and the critical temperature at the junction is $T_c=10.5\pm1K$. For optimally doped sample $x=0.15$ (b):  $\Delta=3.3\pm0.2meV$, $T_c(film)=16.3\pm2K$. And for overdoped sample :$x=0.17$ (c): $\Delta=2.5\pm0.3meV$, $T_c=12.5\pm0.3K$.\label{dIdV}}
\label{dIdVData}
\end{center}
\end{figure*}

In figure \ref{dIdV} we present the typical conductance characteristics for
the three doping levels studied at low temperatures. The
superconducting energy gap $\Delta$ is determined by the energy at which the coherence peaks are observed at low temperatures. The obtained values are consistent with
previous measurements.\cite{DiamantSingle} Additional features
appear at energies higher than $\Delta$.

\par
\begin{figure}
\includegraphics[width=1\hsize]{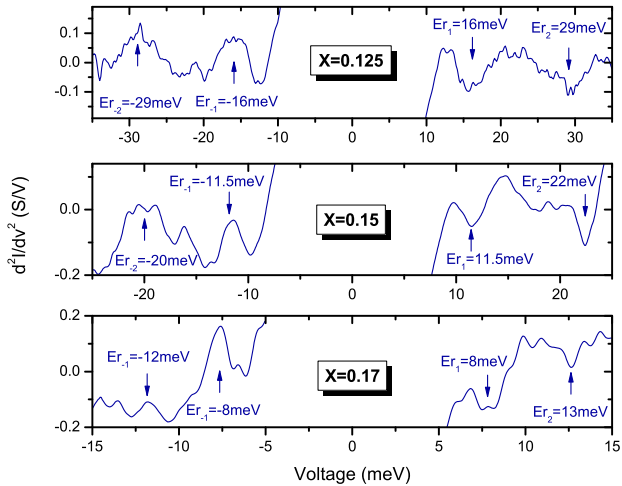}
\caption{(Color online) The derivative of the differential conductance ($d^2I/dV^2$) as a function of the applied bias for the three different doping levels appearing in figure \ref{dIdVData}:  (a) the extremely underdoped $x=0.125$ J0105, (b) the optimally doped $x=0.15$ J0803, and (c) the overdoped $x=0.17$ J0224. For each sample we note the energy at which the bosonic mode appear.\label{d2IdV2}}
\end{figure}

To accurately determine the bias at which the high energy features appear, and to understand their origin, we plot the
derivative of the conductance as a function of voltage (figure \ref{d2IdV2}). The energy of the features are determined from the
position of the step down in the conductance spectra, which manifests itself as a dip in the second derivative for positive biases.\cite{McmillanPhonon, ZhaoFineStructure} A dip in
the positive bias side will appear as a peak in the negative bias
side and vise versa. We ignore features that do not follow this
behavior. Features strictly related to superconductivity will disappear above the critical temperature and critical magnetic field. We were able to verify this on one sample for each doping. 

\par
Below we explain how we differentiated the high energy features from noise and other junction related spurious effects.  First, we made sure
that these features do not arise from the contact itself.
We repeated the measurements for $x=0.15$ and  $x=0.17$ to make sure that the results are junction and film independent. Second, it is possible that geometrical bound
states in the film manifest themselves as a resonance in the
conductance spectra.\cite{NguyenASJBoundStates}. We measured the optimally doped sample at different thickness and found that the high
energy features are thickness independent in contrast with the
expected bound state energy. The third possibility is inelastic
phonon assisted tunneling.\cite{KirtleyInelasticTransport, FranzPhononAssisted} For this case a similar step
in the conductance is expected at the phonon energy. The high
energy features discussed here are not due to boson assisted
tunneling for the following reasons: They disappear below the
critical magnetic field and the critical temperature. While for
the inelastic case one does not expect such strong field and
temperature dependencies. In addition, the case of phonon assisted
tunneling is unlikely due to the strong doping dependence
discussed below. Last,  an additional very large dip feature appears in the $d^{2}I/dV^{2}$ before the first dip we identify as a bosonic mode. This feature is not related to the bosonic modes for the following reasons. The feature appears due to the conductance drop from the coherence peak and does not appear as a step down in the $dI/dV$ spectra. In Addition, it does not have a second harmony (as explained below). Finally, The feature's energy varies  between samples of the same doping level.

\par
We can therefore relate these high energy features to bosonic excitations appearing in the conductance spectra through the frequency dependence of superconducting energy gap $\Delta(\omega)$. Following the logic of Rowell and McMillan we expect the bosonic energy to be shifted by the amplitude of the energy gap. We can therefore
define the excitation energy as $\Omega_{1,2}
=E_{r_{1,2}}-\Delta$, where $E_{r_{1,2}}$ is the energy of the high
energy feature taken from the data, $\Delta$ is the amplitude of the energy gap, and $\Omega_{1,2}$ is
the bosonic resonance energy.
\par
In figure \ref{d2IdV2} we present the differential condactance derivative focusing on the low bias region. In this region the superconducting gap as well as  the bosonic features appear. While $\Delta$ clearly appears in the differential conductance, the bosonic features are conspicuous in the conductance derivative.
For phonons, Morrel and Anderson showed, using the Einstein approximation, that the gap parameter $\Delta(\omega)$ exhibits features at the phonon frequency and its multiplications: $\omega_L$, $2\omega_L$, ... with $\omega_L$ a longitudinal phonon frequency.\cite{MorellAnderson}. Similar results were obtained by Swihart \cite{Swihart} and Culler \cite{Culler} using a Debye approximation. These theoretical findings were also confirmed experimentally.\cite{Giaever1962}
This pattern also appears in our data: $\Omega_2=2\Omega_1$. This strengthens the identification of these features as  bosonic modes related to superconductivity.
\par
In view of the above we made the following steps in order to find the energy of the bosonic mode for each doping level: First, we found the energy of the dip (peak) in the positive (negative) bias and the error as the full width half maximum for the first and second harmonic. We then averaged the positive and negative bias,  $\Omega=\frac{1}{2}(\Omega_1+\Omega_2/2)$, as well as the error. This procedure was carried out for each sample and each junction.

\par
\begin{figure}
\includegraphics[width=1.8\hsize]{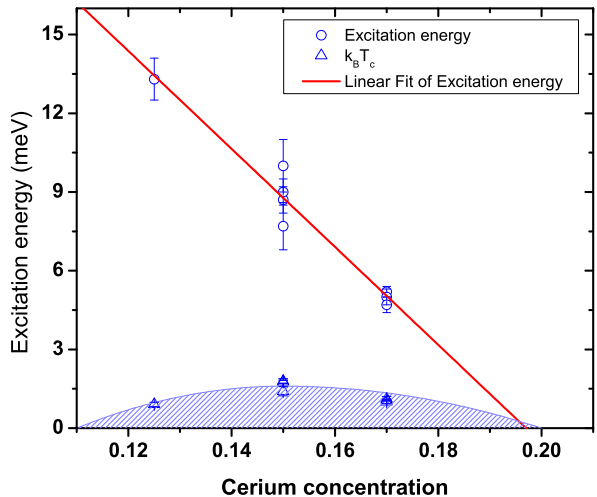}
\caption{(Color online) The spin excitation energy ($\circ$) and the critical temperature ($\triangle$) (The dome is outlined as a guide to the eye) as a function of doping. Each symbol corresponds to a different junction (The error bars reflect the averaged width of both features). We note that the gap amplitude in our measurements follows the critical temperature behavior as a function of doping as reported elsewhere.\cite{DiamantSingle} From our analysis we find that $\Omega$ strongly decreases  with
increasing doping. We draw a linear fit of the excitation energy as a function of the doping, and find that the energy extrapolates to zero at $x=0.2$, which is just beyond the superconducting dome.\label{OmegaVsX}}
\end{figure}

In figure \ref{OmegaVsX} we plot the energy of the bosonic mode, $\Omega$, as a
function of doping. We note that the gap amplitude in our measurements follows the critical temperature behavior as a function of doping as reported elsewhere.\cite{DiamantSingle} From our analysis we find that $\Omega$ strongly decreases  with
increasing doping. For the doping range studied, the amount of cerium substitution changes by merely $1\%$. We therefore do not expect such a strong variation of lattice vibration frequency by such a minute change. Despite the above, a doping driven structural phase transition can manifest itself in an abrupt change in the phonon spectrum. Such a transition, however, is not observed for PCCO in this doping regime.\cite{ArmitageRMP, TarasconStructureNCCO}

\par
We draw a linear fit of the excitation energy as a function of the doping, and find that the energy extrapolates to zero at approximately $x=0.2$. We are not aware of a theoretical prediction for the doping dependence of the bosonic mode. The fact that the line extrapolates to zero right at the edge of the superconducting dome is intriguing. 

\par
We summarize our experimental observations. We observed resonances in the point contact spectra appearing in $\Omega+ \Delta$, and $2\Omega+ \Delta$. These features disappear above the critical temperature, and above the critical magnetic field.  We relate them to bosonic excitations responsible for superconductivity. We find that these bosonic excitations are doping dependent. 

\par
We shall now discuss our results in the context of other experiments and predictions.  Clear tunneling features were observed in previous planar tunneling experiments \cite{DiamantSingle,DirtySC}, however, these measurements did not exhibit conspicuous signatures of bosonic structures in the conductance spectra. A possible explanation for this could arise from the large contact size used relative to the coherence length, $\xi$. This can be understood noting that STM measurements on the hole-doped Bi$_2$Sr$_2$CaCu$_2$O$_{8-\delta}$ showed large spatial variation in the gap amplitude over a length scale of the order of $\xi$.\cite{LeeBosonicMode, LangGranular} For the electron-doped cuprates $\xi$ is ten times larger than for Bi$_2$Sr$_2$CaCu$_2$O$_{8-\delta}$,\cite{HowaldBSCCOCoherence} yet, it is much smaller comparing to the large scale of the planar junctions. Since the bosonic excitation is shifted by the value of the gap, \cite{HarrisonPhonon} and since tunneling in this large scale contact is actually a measurement of an averaged gap amplitudes, the bosonic excitation energy is smeared on the energy axis. In contrast, the junction size in point contact measurements is much smaller. It is of the order of the coherence length of PCCO. It is therefore possible that this point contact junction size is sufficient to obtain a relatively constant gap amplitude and consequently to enable the observation of bosonic modes. 

\par
We compare our result to the phenomenological relation $\Omega=5.8k_{B}T_{c}$ drawn from neutron scattering experiments.\cite{WilsonNeutron} ,For the optimally doped sample we get $\Omega=8.2meV$, which falls within the error bar of our results.  For the overdoped sample this relation falls slightly above our data point. However, the underdoped sample deviates from this relation. We speculate that this difference stems from the extent of the \AFM~ region in the n-doped cuprates compare to the p-doped ones.

\par
Finally, Niestemski \etal\cite{NiestemskiBosonicMode} analyzed STM data to identify bosonic excitations in optimally doped PLCCO. However, Guo-meng Zhao \cite{ZhaoFineStructure} noted, based on the analysis of McMilan and Rowell \cite{McmillanPhonon}, that the bosonic energy should be identified as a step down in the conductance which yields a dip in the second derivative for positive biases. This prediction also appears in the theoretical work by Ar. Abanov and A. V. Chubukov \cite{AbanovChubokovSIN}, in their work they also predict that the bosonic energy, $\Omega$, should be smaller than twice the energy gap. This is in contrast to our data. 

In summary, we observed structures in the derivative of the conductance characteristics. We eliminate boson assisted tunneling, spurious noise and geometrical effects, as possible explanations for these features. The temperature, field and voltage dependencies of these features suggest that they arise from bosonic modes reflected through the energy dependence of the superconducting gap. Pairing by both phonons and magnetic excitations predicts a step down feature in the tunneling spectra. However, the doping dependence of the boson energy casts strong doubts on phonon mediated superconductivity in the electron-doped cuprates. Our results are therefore more in line with the spin excitation paring theory.

\begin{acknowledgments}
We thank A. Yechzkel for machining the differential screw.
This research was partially supported by the Binational Science
Foundation grant number 2006385, and the Israel Science Foundation
grant number 1421/08.
\end{acknowledgments}

\bibliographystyle{apsrev}
\bibliography{myBib}
\end{document}